\documentclass{sig-alternate-05-2015}
\usepackage{balance}  
\usepackage{graphics} 
\usepackage{epstopdf}
\usepackage{times}    
\usepackage{latexsym}
\usepackage[bf,small]{caption}
\usepackage{subfig}
\usepackage{wrapfig}
\usepackage{enumitem}
\usepackage{epsfig}

\usepackage{ucs}
\usepackage[utf8]{inputenc}
\usepackage[english]{babel}
\usepackage{fontenc}
\usepackage{graphicx}

\usepackage[dvips]{hyperref}

\date{05/10/2016}
\makeatletter
\let\@copyrightspace\relax
\makeatother

\begin{document}
                    
\title{Summarizing Situational and Topical Information \\During Crises}

\numberofauthors{6} 
%
\author{
%
%
\alignauthor
Koustav Rudra\\
       \affaddr{IIT Kharagpur, India\\
       koustav.rudra@cse.iitkgp.ernet.in}\\
\alignauthor
Siddhartha Banerjee\\
       \affaddr{The Pennsylvania State University, USA\\
       sub253@ist.psu.edu}\\
\alignauthor 
Niloy Ganguly\\
       \affaddr{IIT Kharagpur, India\\
       niloy@cse.iitkgp.ernet.in}\\
\and  
\alignauthor 
Pawan Goyal\\
       \affaddr{IIT Kharagpur, India\\
       pawang.iitk@gmail.com}
\alignauthor 
Muhammad Imran\\
\affaddr{Qatar Computing Research Institute, HBKU, Doha, Qatar\\
	mimran@qf.org.qa}
\alignauthor Prasenjit Mitra\\
\affaddr{The Pennsylvania State University, USA\thanks{This work was done when the author was at Qatar Computing Research Institute, HBKU, Doha, Qatar.}\\
	pmitra@ist.psu.edu}
}

\maketitle

\begin{abstract}

The use of microblogging platforms such as Twitter during crises has become widespread. More importantly, information disseminated by affected people contains useful information like reports of missing and found 
people, requests for urgent needs etc. For rapid crisis response, humanitarian organizations look for situational awareness information to understand and assess the severity of the crisis.
In this paper, we present a novel framework (i) to generate abstractive summaries useful for situational awareness, and (ii) to capture sub-topics and present a short informative summary for each of these topics.
A summary is generated using a two stage framework that first extracts a set of important tweets from the whole set of information through an Integer-linear programming (ILP) based optimization technique and then 
follows a word graph and concept event based abstractive summarization technique to produce the final summary.
High accuracies obtained for all the tasks show the effectiveness of the proposed framework.

\end{abstract}



\vspace{2mm}

\section{Introduction}
\label{sec:intro}

Microblogging platforms such as Twitter provide rapid access to situation-sensitive information that people post during mass convergence events such as natural disasters. Rapid crisis response can be aided by 
processing these tweets~\cite{vieweg2014integrating} in real-time. Different stake-holders (e.g. different humanitarian organizations) have different informational needs. For example, to better understand the 
severity and status of an event, most of these organizations rely on situational awareness information. Some others look for information on a specific concern like reports of damage to key infrastructure in the 
area such as airports, bridges, buildings, communication infrastructure, etc.

Typically, the first step in extracting situational awareness information from these tweets involves classifying them into different informational categories such as infrastructure damage, shelter needs or offers, 
relief supplies, etc. For instance, one such application, AIDR~\cite{152014aidr}, classifies Twitter messages into different categories in real time. However, even after the automatic classification step, 
each category still contains thousands of messages many of which are important.  Additional in-depth analysis is required to create a coherent situational awareness summary for disaster managers to understand 
the situation, which can be rapidly changing.

In this paper, we seek to extract important topical information from microblogging platforms and generate summaries for the identified topics.
For example, within the tweets categorized as infrastructure damage related, the reader can examine the status of airports, buildings, bridges, etc. provided this information has been reported. 
Drilling down into sub-topics and examine the set of tweets from which the information was extracted, we must group tweets dealing with similar information into sets. These sets should be labeled 
with a concept name.  

~\\
\noindent{\bf Summarizing Messages in Disaster-related Categories:} Summarizing tweets is significantly more challenging than 
summarizing news articles. The difficulty arises because tweets are often written in informal and non-standard language as opposed to the formal language used in news articles. 
To address the real-time nature of our application and the need for a more readable, more informative, and more easily understandable summary, we propose a novel {\it two-step summarization process} 
that uses a fast extractive summarization technique~\cite{gupta2010survey} followed by an abstractive summarization step that improves the information coverage and readability of the final summary. 
Rudra~et~al. used extractive summarization to summarize a set of tweets~\cite{rudra-cikm-2015}. For example, consider the following tweets collected during the Nepal earthquake in 2015:

\begin{enumerate}
\scriptsize
 \item {\tt Tribhuvan international airport closed after the quake}
 \item {\tt Airport closed after 7.9 Earthquake in Kathmandu}
\end{enumerate}
An abstractive summary of these tweets would be as follows: \\
{\tt Tribhuvan international airport closed\\ after 7.9 earthquake in Kathmandu}.  Note that the latter is more compact freeing up words that can be used for additional information coverage. 

Information coverage can be improved in a summary 
by including as many  content words as possible. 
For example, Rudra~et~al. have showed that maximizing the coverage of content words
produces effective summaries
of disasters~\cite{rudra-cikm-2015}. However, we observed that many such content words are semantically
similar and capturing one of those in final summary will suffice to 
provide adequate information coverage. Hence, in this work,
we collate similar nouns and verbs to
develop concept and event clusters.
We propose a word-graph based abstractive summarization technique 
that combines information from semantically similar tweets (extracted in first step) and 
applies an ILP-based\footnote{Henceforth we represent integer linear programming approach as ILP-based approach} content word (numeral, location, concept, event) coverage method to generate the summary. 

Although abstractive summarization~\cite{olariu-2014} produces more compact and informative sentences, the algorithms in general are time-consuming. Hence, if the abstractive approach is run over 
the entire incoming set of tweets, it may not be possible to produce the results in real-time (which is one of the important requirements during disasters). 
In order to circumvent this problem, first we extract a set of important tweets from the whole set
using a fast but effective extractive summarization approach. 
In the second step, we use
abstractive summarization to choose and rewrite the most important tweets among them, 
remove redundancy and improve the readability of the tweets.
\\

\noindent{\bf Identification and Summarization of Micro-topics:} To provide information about
events at a finer granularity such as when an airport has been shut, or re-opened, school suspended, communication cut or restored etc. that happen during a crisis situation, our method first 
identifies micro-topics (i.e., small-scale events) and then generates summaries for each of these micro-topics at a 
finer granularity level.

In this work, we use the Nepal earthquake dataset~\cite{nepal-quake-wiki} comprising of several million tweets collected and initially classified by the AIDR platform~\cite{152014aidr}.
Our contribution lies in the two-step extractive-abstractive summarization approach (Section \ref{sec:summarize}) that is efficient
and yet generates better summaries with respect to information coverage, diversity, coherence, and readability. 
 Experimental results in Section~\ref{sec:topic_summarization} also confirm that 
our extracted topics and summaries related to those topics outperform traditional LDA based methodologies.
Finally, we conclude our paper in Section~\ref{sec:conclu}.

\section{Related work}
\label{sec:related}
Real-time information posted by affected people on Twitter helps improve disaster relief operations~\cite{gao2011harnessing,imran2015processing}. However, 
relief organizations can plan more effectively if they have access to 
crucial information from the tweets~\cite{klein2012detection}. 

Kedzie et al.~\cite{kedzie-acl-2015} proposed an {extractive summarization} method to summarize disaster event-specific information from news articles. In contrast, several researchers have attempted to 
utilize information from Twitter to retrieve important situational updates from millions of posts on  disaster-specific events~\cite{verma-nlp-situation-awareness}.
More recently, sophisticated methods for automatically generating summaries by extracting the most important tweets on the event~\cite{rudra-cikm-2015} have been proposed. To generate summaries in real-time, 
a few approaches for online summarization of tweet streams have recently been proposed~\cite{rudra-cikm-2015}.  

The methods mentioned above generate extractive summaries that are merely a collection of tweets. 
Ideally, we prefer an abstractive summary composed of
important content from tweets instead of the whole tweets.
Such a summary should also be more readable than a collection of tweets. Furthermore, the summaries should not contain redundant information. 
To this end, Olariu~\cite{olariu-2014} proposed a bigram word-graph-based summarization technique,
which is capable of handling online streams of tweets in real-time 
and also generates abstractive summaries. Each bigram represents a node in the graph and new words are added in 
real-time from incoming new tweets.
However, the method does not consider POS-tag information of nodes and thus can spuriously fuse tweets having the same bigram but are otherwise unrelated.
Furthermore, it is a general method that
does not consider the typicality of disaster 
related tweets, for example earlier Rudra~et~al~\cite{rudra-cikm-2015} showed that during disasters content words~(nouns, verbs, numerals) vary quite slowly compared to any other general events like sports, 
movies etc. In our proposed abstractive summarization framework, we have incorporated such domain dependent features to make the summary more coherent, informative, and useful.
Banerjee, Mitra, and Sugiyama proposed a graph-based abstractive summarization method on news articles~\cite{banerjeemulti}.  
Several new sentences are generated using
a graph where words are nodes, edges are added between two consecutive words present in a sentence and an optimization problem is formulated that selects the best sentences from the new sentences to optimize the overall quality of the summary. The optimization problem ensures that redundant information is not conveyed in the final generated summary. However, the graph construction and path generation is computationally expensive and cannot be used in real-time.

We combine the positive aspects of the above studies - (a) we employ extractive summarization to reduce the number of tweets, and on the reduced set run an algorithm adapated from
the technique proposed by Banerjee~et~al.~\cite{banerjeemulti} for tweet fusion 
(b) we use POS tags along with the words in each bigram to avoid spurious tweet fusions and (c) 
we employ disaster-specific content words to determine the importance of a disaster-related tweet \cite{rudra-cikm-2015}. 
Further, we also focus on template-based topic extraction and summarizing information over those topics.

\section{Dataset and Classification of Messages}
\label{sec:dataset}
\noindent We use the Nepal Earthquake 2015 Twitter data from CrisisNLP~\cite{crisisNLP2016}.
The dataset consists of 27 million messages from April 25th to April 27th obtained using different keywords 
(e.g. Nepal Earthquake, NepalQuake, NepalQuakeRelief, NepalEarthquake, KathmanduQuake, QuakeNepal, EarthquakeNepal, $\cdots$, etc.). 

In this work, we selected AIDR~\cite{152014aidr} classified messages from three categories for which the machine confidence was $\geq$ 0.80. The selected classes and messages in each of the three classes 
are as follows:

1. {\bf Missing, trapped, or found people:} 10,751 2. {\bf Infrastructure and utilities:} 16,842 3. {\bf Shelter and supplies:} 19,006 messages.

\section{Automatic Summarization}
\label{sec:summarize}
\noindent Given the machine-categorized messages by AIDR, in this section we present our two step automatic summarization approach to generate summaries from each class.  
We consider the following key characteristics/objectives while developing an automatic summarization approach:

\begin{enumerate}[leftmargin=*]
 \item A summary should be able to capture most situational updates from the underlying data. That is, the summary should be rich in terms of information coverage.

 \item As most of the messages on Twitter contain duplicate information, we aim to produce summaries with less redundancy while keeping important updates of a story.
 
 \item Twitter messages are often noisy, informal, and full of grammatical mistakes. We aim to produce more readable summaries as compared to the raw tweets.

 \item The system should be able to generate the summary in real-time, i.e., the system should not be heavily overloaded with computations such that by the time the summary is produced, 
 the utility of that information is marginal.
 
\end{enumerate}

The first three objectives can be achieved through abstractive summarization and near-duplicate detection, 
however, it is very difficult to achieve that in real-time (hence violating the fourth constraint). In order to fulfill these objectives, 
we follow an extractive-abstractive framework to generate summaries. We define our overall
summarization framework as CONcept based ABstractive Summarization ({\bf CONABS}).
In the first phase (extractive phase), we use the approach proposed by Rudra et al.~\cite{rudra-cikm-2015} and 
select a subset of tweets that cover most of the information produced and then run abstractive summarization over that.
We generate the paths using the extractive-abstractive framework proposed by Rudra et al~\cite{rudra2016summarizing}.
Our goal is to select the best paths from these generated tweet paths with the objective of generating a readable and 
informative summary. To this end, we formulate an ILP based technique that selects final paths and generates the summary.

~\\
\noindent {\bf Concept and event extraction: }
\label{sec:concept_extraction}
Given that AIDR classified messages into the categories mentioned above, we extract important concepts and events associated with them. For example, the `infrastructure' class contains information about 
building collapse, temples and whether the airport is open or closed.
We observed that such micro-level information mainly consists of two core nuggets, a noun part which we call as a concept (e.g., airport) and a verb part, which we call as an event (e.g., closed).
In our summarization process, we capture information about these concepts by using nouns, because concepts are in general denoted by the nouns~\cite{li2014empirical}. For this purpose, (i)~extract all the nouns
from the dataset, (ii)~develop a complete undirected weighted graph where nouns are nodes and weight of the edge between two nouns is their semantic similarity score (we have used
lin similarity measure), and (iii)~run affinity clustering method to cluster semantically
similar nouns (e.g., airport, flight). Each of the identified clusters represents a particular concept.

~\\
Ritter~et~al~\cite{Ritter12} proposed 
a method to extract events from tweets but this method takes significant amount of time to tag large stream of tweets. This creates a bottleneck in real-time summarization process.
Hence, we cannot use their method directly in our proposed summarization approach.
We observed that main verbs generally represent such
events like `collapsed', `killed', `injured', `blocked'.
We construct a complete undirected weighted graph by taking the verbs and apply clustering technique over the graph (similar to concept extraction).
Each cluster of verbs represents one event. For example, verbs like `injured', `wounded' are clubbed into one cluster and represent one event.

~\\
\noindent {\bf ILP Formulation}
~\\
For abstractive summarization phase, we redefine the content words. Content words consist of numerals, places (this is similar to that adopted during the 
extractive phase), concepts, and events.    
The ILP-based technique optimizes based upon three factors - (i)~Presence of content words: The formulation tries to maximize the number of these parameters in the final summary which
in turn takes care of diversity by reducing the probability of choosing the same content word multiple times.
(ii)~Informativeness of a path, i.e., finding importance of a path based on centroid-based ranking~\cite{radev2004centroid}, 
and (iii)~{\it Linguistic Quality Score} that captures 
the readability of a path using a trigram confidence score~\cite{heafield2011kenlm}.

\begin{table}[t]
\center 
\caption{{\bf Notations used in the summarization technique}}
\resizebox{\columnwidth}{!}{
\begin{tabular}{|p{0.18\columnwidth}|p{0.75\columnwidth}|}
\hline
{\bf Notation} & {\bf Meaning} \\
\hline
$L$ & Desired summary length (number of words) \\
\hline
$n$ & Number of {\em tweet-paths} considered for summarization (in the time window specified by user)\\
\hline
$m$ & Number of distinct content words included in the $n$ {\em tweet-paths} \\
\hline
$i$ & index for {\em tweet-paths} \\
\hline
$j$ & index for content words \\
\hline
$x_{i}$ & indicator variable for {\em tweet-path} $i$ (1 if {\em tweet-path} $i$ should be included in summary, 0 otherwise)\\
\hline
$y_{j}$ & indicator variable for content word $j$ \\
\hline
$Length(i)$ & number of words present in {\em tweet-path} $i$\\
\hline
I($i$) & Informativeness score of the {\em tweet-path} $i$\\
\hline
LQ($i$) & Linguistic quality score of a {\em tweet-path}\\
\hline
$T_{j}$ & set of {\em tweet-paths} where content word $j$ is present\\
\hline
$C_{i}$ & set of content words present in {\em tweet-path} $i$\\
\hline
\end{tabular}}
\label{table:ILP-parameters}
\vspace*{-3mm}
\end{table}

The summarization of $L$ words  is achieved by optimizing the following ILP objective function, whereby the highest scoring {\em tweet-paths} are
returned as output of summarization, The equations are as follow:
\begin{equation}
max( \sum_{i=1}^{n}LQ(i).I(i).x_{i} \, + \, \sum_{j=1}^{m} y_{j})	
\label{eqn:optimize}
\end{equation}
subject to the constraints
\begin{equation}
\label{eqn:length_constraint}
\sum_{i=1}^{n} x_{i} \cdot Length(i) \leq L \\
\end{equation}
\begin{equation}
\label{eqn:content-word-constraint}
\sum_{i\in T_{j} } x_{i} \geq y_{j}, j=[1 \cdots m] \\
\end{equation}
\begin{equation}
\label{eqn:tweet_constraint}
\sum_{j\in C_{i}} y_{j} \geq |C_{i}| \times  x_{i}, i=[1 \cdots n]
\end{equation}
where the symbols are as explained in Table~\ref{table:ILP-parameters}.
The objective function considers both the number of {\em tweet-paths} included in the summary
(through the $x_i$ variables) as well as the number of important content-words 
(through the $y_j$ variables) included.
The constraint in Eqn.~\ref{eqn:length_constraint}
ensures that the total number of words contained in the {\em tweet-paths} that get included in the summary is at most
the desired length $L$ (user-specified)
while the constraint in Eqn.~\ref{eqn:content-word-constraint}
ensures that if the content word $j$ is selected to be included in the summary,
i.e., if $y_{j} = 1$,  
then at least one {\em tweet-path} in which this content word is present is selected. 
Similarly, the constraint in Eqn.~\ref{eqn:tweet_constraint} ensures that
if a particular {\em tweet-path} is selected to be included in the summary, then the content words
in that  {\em tweet-path} are also selected. 

We use the GUROBI Optimizer~\cite{gurobi}
to solve the ILP. After solving this ILP, the set of  {\em tweet-paths} $i$ such that $x_{i}=1$, represent the summary at the current time.

\subsection{Experimental Setup and Results}
\label{sec:experiment}

In this section, we compare the performance of our proposed framework with state-of-the-art abstractive and disaster-specific summarization techniques. 
Given the AIDR classified messages from three classes, we perform date-wise split starting from 25th April to 27th April.

\noindent{\bf Baseline approaches: }
We use three state-of-the-art summarization approaches described below:
\begin{enumerate}[leftmargin=*]
 \item {\bf COWTS:} an extractive summarization approach specifically designed for generating summaries from disaster-related tweets~\cite{rudra-cikm-2015}. 
 \item {\bf APSAL:} an affinity clustering based summarization technique proposed by Kedzie et al.~\cite{kedzie-acl-2015}. It mainly considers news articles and focuses on human-generated information 
 nuggets to assign salience score to those news articles while generating summaries.
 \item {\bf TOWGS:} an online abstractive summarization approach proposed by Olariu~\cite{olariu-2014}. It is designed for informal texts like tweets. They consider bigrams as nodes and build word graph 
 using these nodes. To generate a summary, they start from most frequent bigrams to explore different paths.
 In our case, we modified it to generate event-specific summaries as it was originally not proposed to do so.
\end{enumerate}

\noindent{\bf Evaluation using expert generated data}
We took summaries generated by experts from the disaster management domain. During Nepal earthquake, UN OCHA (United Nations Office for the Coordination of Humanitarian Affairs) among other 
humanitarian organizations used AIDR's output (i.e., machine classified messages) for their disaster response efforts. In this case, the experts were given the machine classified messages that they analyzed to
generate a situational awareness report for each informational category. We consider these reports as our gold standard summaries.

\begin{figure}[tb]
\centering
\fbox{\includegraphics[width=0.95\columnwidth,height = 4cm]{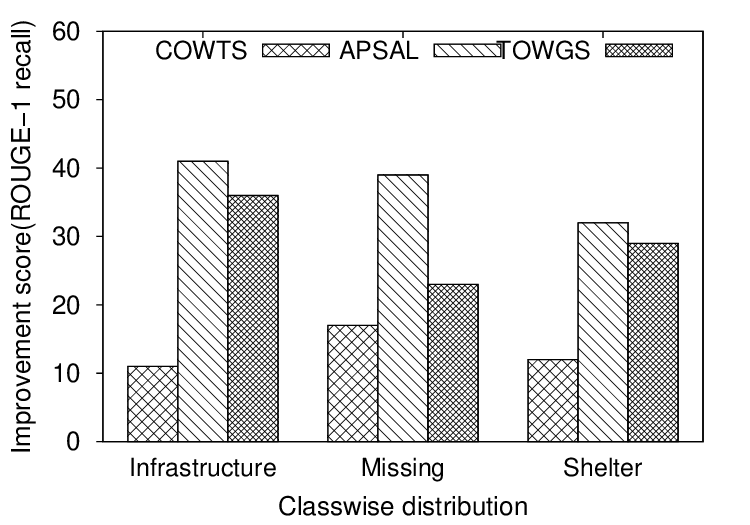}}
\caption{{\bf Improvement in ROUGE-1 recall score (averaging over three days) by CONABS over baseline methods (COWTS, APSAL, TOWGS)}}
\label{fig:rouge_score}
\vspace{-4mm}
\end{figure}

Figure~\ref{fig:rouge_score} shows the improvement by our method over baseline techniques in terms of ROUGE-1 recall score which basically indicates in percentage the amount of more (important) information 
covered in the generated summaries. We can see that CONABS performs significantly better compared to other three baselines - the improvement ranges from 10\% to 40\%.
~\\
\noindent{\bf Evaluation using crowdsourcing:} We perform crowdsourced evaluation using the CrowdFlower~\footnote{\url{http://www.crowdflower.com/}} crowdsourcing platform. 
We take summaries generated from each class using our proposed method and all three baselines for each day---in total we use 9 summaries. 
A crowdsourcing task, in this case, consists of four summaries (i.e., one proposed and three from baseline methods) and the four criteria with their description (as described below) along with a scale 
from 1 (very bad) to 5 (very good) for each criterion. For each task, we asked five different annotators to read each summary carefully and provide scores for each criterion. The exact description of the 
crowdsourcing task is as follows: 

\textit{``The purpose of this task is to evaluate machine-generate summaries using tweets collected during the Nepal Earthquake happened in 2015. Each task given below has 4 summaries of length 200 words 
generated by 4 different algorithms on the same set of tweets (thousands in this case) belonging to a particular topic.  Given the summaries and their topic, we are interested in comparing them based on the 
following criteria: Information coverage, Diversity and Readability''}. 

The definitions of various criteria we used in the task and discussion of the results are as follows:
~\\

\begin{figure}[h]
\vspace{-4mm}
\center
\subfloat[{25th (info. coverage)}]{\includegraphics[width=0.5\columnwidth]{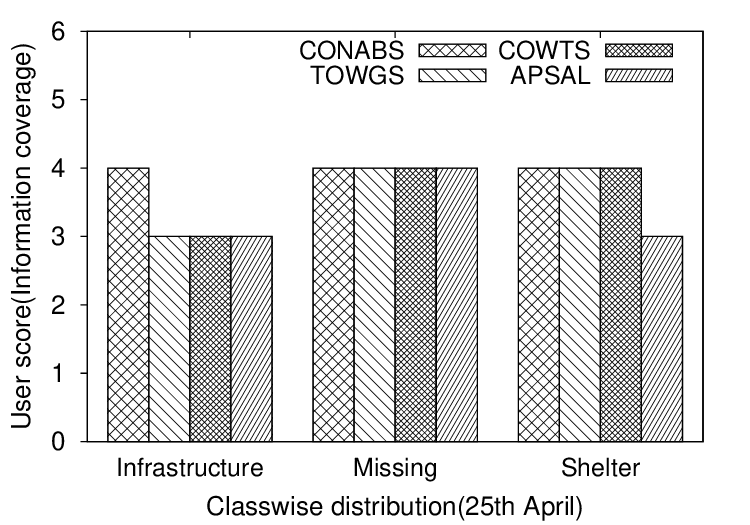}}
\hfil
\subfloat[{26th (info. coverage)}]{\includegraphics[width=0.5\columnwidth]{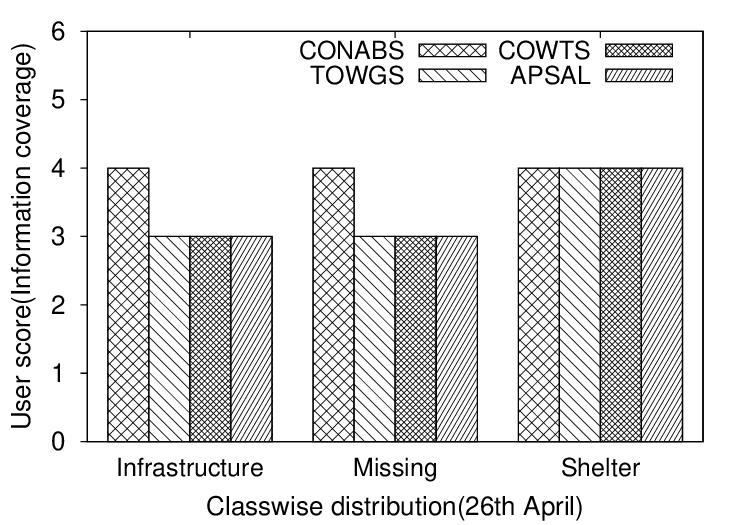}}
\caption{{\bf Results of the crowdsourcing based evaluation based on the information coverage}}
\label{fig:crowd_info_cover}
\vspace{-3mm}
\end{figure}

\noindent{\bf Information coverage} corresponds to the richness of information a summary contains. For instance, a summary with more informative 
sentences (i.e., crisis-related information) is considered better in terms of information coverage. Our proposed method
is able to capture very good situational information/updates in case of Infrastructure and Missing classes for both of the days chosen while
it performs fairly in the shelter class. In 4 cases, it performs better than the three competing techniques, and it performs equally well with COWTS, TOWGS, and APSAL in 3, 1, 1 cases respectively.
Figure~\ref{fig:crowd_info_cover} shows the detailed ratings of users for 25th and 26th April~\footnote{We only keep two dates to maintain clarity and brevity}.

\begin{figure}[h]
\vspace{-4mm}
\center
\subfloat[{25th (info. diversity)}]{\includegraphics[width=0.5\columnwidth]{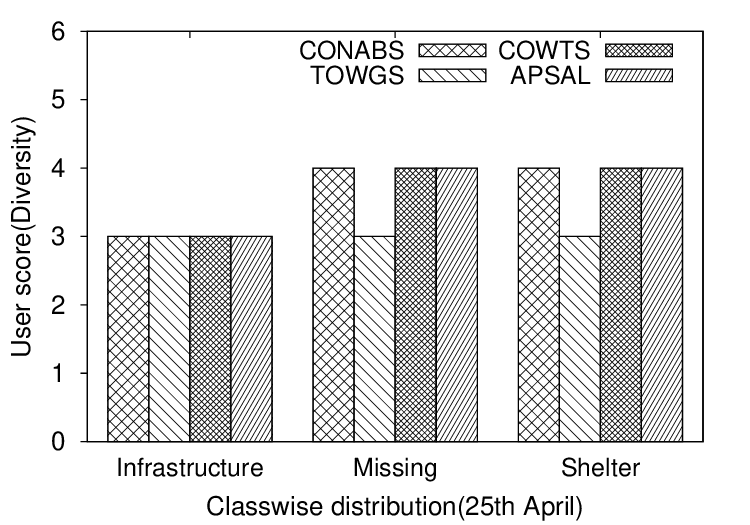}}
\hfil
\subfloat[{26th (info. diversity)}]{\includegraphics[width=0.5\columnwidth]{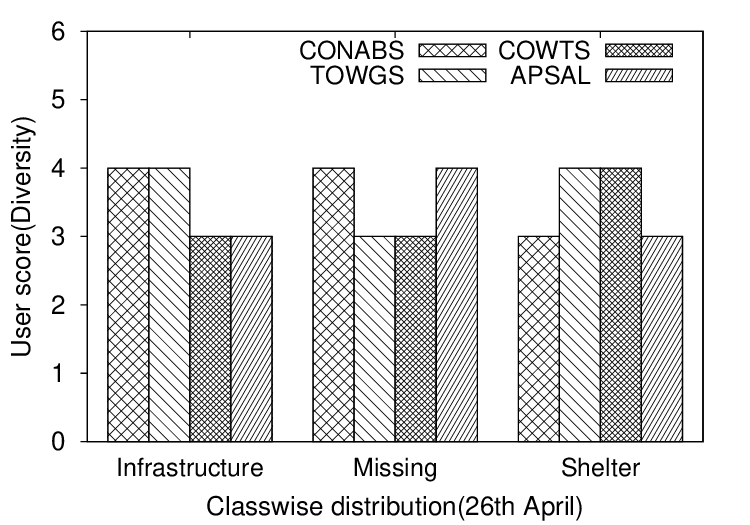}}
\caption{{\bf Results of the crowdsourcing based evaluation based on the information diversity}}
\label{fig:crowd_diversity}
\vspace{-3mm}
\end{figure}

\noindent{\bf Diversity} corresponds to the novelty of sentences in a summary. A good summary should contain diverse informative sentences. While we do not apply
any direct parameter in our ILP framework to control diversity, in our abstractive ILP method, we not only rely on the importance score of paths but also coverage of
different content words, which helps in capturing information from various dimensions. This is also quite clear from Figure~\ref{fig:crowd_diversity}. 
In seven out of nine cases, CONABS generated summaries that are comparable to other baseline techniques.

\begin{figure}[htb]
\vspace{-4mm}
\center
\subfloat[{25th (readability)}]{\includegraphics[width=0.5\columnwidth]{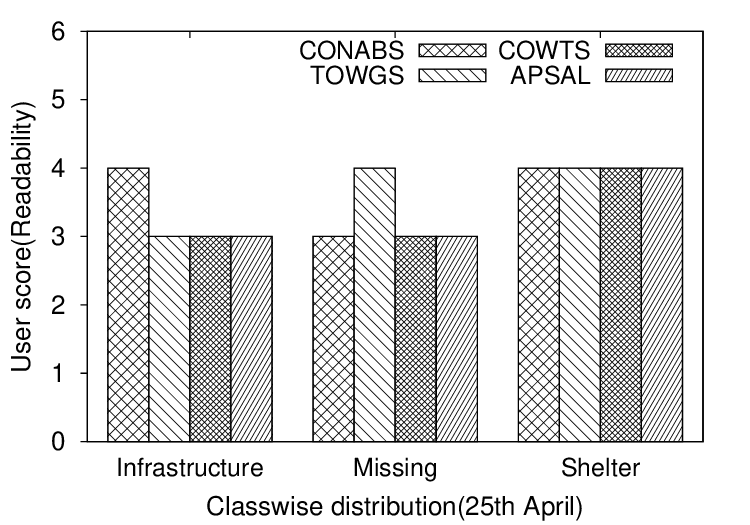}}
\hfil
\subfloat[{26th (readability)}]{\includegraphics[width=0.5\columnwidth]{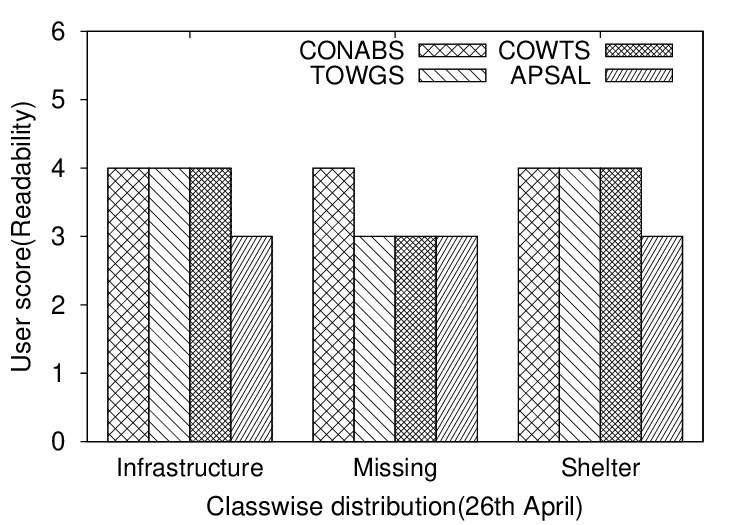}}
\caption{{\bf Results of the crowdsourcing based evaluation based on the readability}}
\label{fig:crowd_readability}
\vspace*{-3mm}
\end{figure}

\noindent{\bf Readability} measures how easy it is to read 
the summary. A good summary should be easily readable, well formed, coherent, and have fewer grammatical errors.
We used a linguistic quality score in our final ILP framework
to generate coherent summaries.
Our system chooses paths with higher linguistic scores.
Summaries generated by CONABS were rated to be equal or better than the
other baselines in 8 (of nine) cases. Figure~\ref{fig:crowd_readability} shows that
our summaries' lowest readability score was 3. Its performance is particularly good on 26th April where it is marked 4 (good) for all cases. 

CONABS performs as well or better than other baseline techniques in most of the cases. 

\begin{table*}[!htbp]
\center
\caption{{\bf Summary of length 50 words(excluding \#,@,RT,URLs), generated from the situational tweets of the infrastructure class (26th April) by (i)~CONABS (proposed methodology),
(ii)~COWTS.}}
\resizebox{\textwidth}{!}{
\begin{tabular}{|p{0.5\textwidth}||p{0.5\textwidth}|}
\hline 
{\bf Summary by CONABS} & {\bf Summary by COWTS}\\
\hline
Times of india live blog earthquake in katmandu , 25 04 2015. Chairs follow-up meeting to review situation following earthquake in decades. 5 commercial flights have landed in kathmandu was painted in 1850 ad. Iaf's c-130j aircraft carrying 55 passengers , including four infants , lands at delhi's palam airport. Nepal quake photos show historic buildings reduced to rubble as survivor search continues. & \#PM chairs follow-up meeting to review situation following \#earthquake in \#Nepal @PMOlndia  \#nepalquake. @SushmaSwaraj @MEAcontrolroom  Plz open help desk at kathmandu airport.  @Suvasit thanks for airport update. \#NepalQuake. Pakistan Army Rescue Team comprising doctors, engineers \& rescue workers shortly after arrival at \#Kathmandu Airport http://t.co/6Cf8bgeort. RT @cnnbrk: Nepal quake photos show historic buildings reduced to rubble as survivor search continues. http://t.co/idVakR2QOT http://t.co/Z. \\
\hline
\end{tabular}}
\label{table:summary_sample}
\end{table*}

Table~\ref{table:summary_sample} shows summaries generated by CONABS and COWTS (both disaster-specific methodologies) from the same set of messages (i.e tweets form infrastructure class posted on 26th April). 
The two summaries are quite distinct. We find that summary returned by COWABS is more informative and diverse in nature compared to COWTS. For instance, we can see the COWABS summary contains information about 
flights, damages of buildings, and information sources.

In this approach, we have measured semantic LIN similarity based on wordnet for nouns and verbs. However, we observe that in case of verbs this similarity metric does not perform well. As a result, some unrelated
verbs may be clustered and some important information may be missed in final summary. In future, we try to use better semantic similarity measures to resolve this problem.

~\\
\noindent{\bf Time taken for summarization: } {As stated earlier, one of our primary objectives
is to generate the summaries in real time. Hence, we analyze the execution times of the various
techniques. For infrastructure, missing, and shelter classes, our proposed method CONABS takes 25.947, 17.915, and 26.663 seconds on average (over three days) respectively, to generate the summaries. 
The time taken by CONABS is comparable with other real time summarization methods like COWTS, and TOWGS. However, APSAL requires more time due to large nonnegative matrix factorization and computation of 
large similarity matrices.}

\section{Identification of sub-topics and summarization}
\label{sec:topic_summarization}
Following a major crisis, a number of small-scale sub-events such as `power outage', `bridge closure' etc. happen. Normal LDA based topic detection techniques do not capture micro-level sub-events. 
Moreover, according to UN OCHA, such LDA topics are too general to act upon~\cite{vieweg2014integrating}. In this section, we capture sub-events/topics from messages classified in a particular category 
(e.g. infrastructure damage). We define a sub-topic as a combination of a noun and a verb where noun represents a concept and verb represents an event (as described in the previous section). However, in 
this section we seek dependency relations between nouns and verbs, which is important to declare some information as an event/topic. Table~\ref{table:topic_phrase} provides examples of some sub-topic phrases 
from various AIDR classes. These sub-topics show important yet very specific events after the major earthquake crisis. For example, these include `shut down of airports', `resume of flight operation', 
emergency declared etc.

Most of the existing topic detection methodologies represent topics as a bag of words. In our case we try to capture the semantics between nouns and events using dependency relationships. For 
instance, in Table~\ref{table:topic_phrase}, `flight' is related to `shut', and `road' is related to `crack'. To the best of our knowledge, no prior work on processing tweet streams during disasters has 
attempted to combine nouns and events to generate such micro-level topic phrases.

\begin{table}[!htbp]
\center
\caption{{\bf Popular topic phrases posted on the first day of the Nepal earthquake (Apr 25, 2015)}}
\resizebox{\columnwidth}{!}{
\begin{tabular}{|p{0.3\columnwidth}|p{0.7\columnwidth}|}
\hline
{\bf Class} & {\bf Topic phrases} \\
\hline
Infrastructure & `service affect', `shut flight', `crack road', `water report', `topple  tower'\\
\hline
Injured & `casualty grow', `victim treat', `hospital accommodate', `man trap', `casualty injure'\\
\hline
Missing &  `family stuck', `tourist strand', `rescue location', `database track', `contact number'\\
\hline
Shelter & `field clean', `water equip', `emergency declare', `deploy transport', `deploy aircraft'\\
\hline
\end{tabular}}
\label{table:topic_phrase}
\vspace{-2mm}
\end{table}

\noindent {\bf Assigning nouns to events: } In the sub-topic extraction methodology, we have found that it is often non-trivial to associate nouns to the context of an event in a tweet. For example, 
the words `says' and `toppled' in the sentence `\#China media says buildings toppled in \#Tibet http://t.co/O7VSYWTGsk'
were identified as events~\cite{Ritter12}. 
The noun `building' is related to 
the term `toppled' but it is not related to 
the verb `says'. Hence,
(`building',`toppled') forms a valid topic phrase whereas (`building',`says') is not a topic phrase. It is observed that sometimes such nouns may not always appear prior or adjacent to the events in a tweet. 
For example, in `India sent 4 Ton relief material, Team of doctors to Nepal',
(`relief',`sent') forms a valid topic phrase but the noun `relief' appears after the event `sent'. 

If a noun is directly associated/connected with 
an event (edge exists between noun and event in dependency tree), we associate that noun with the
event. We use POS tagger~\cite{postag-2012}, event detector~\cite{Ritter12}, and dependency parser~\cite{twitter_parser} for tweets to extract the association information.

~\\
\noindent {\bf Ranking topic phrases: } In this part, we rank the identified topic phrases. We only keep those topic phrases for which its constituent noun and event occur more than a certain 
threshold value --- in this case we set it as 10. Next, we compute Szymkiewicz-Simpson overlap score between noun (N) and event (E) as follows:
\begin{equation}
\label{eqn:topic_rank_eqn}
Overlap(N,E) = \frac{|X \cap Y|}{min(|X|,|Y|)} 
\end{equation}
where X indicates the set of tweets containing N and Y indicates the set of tweets containing E. Finally, we rank the topic phrases based on the similarity scores computed as per Equation~\ref{eqn:topic_rank_eqn}.

~\\
\noindent {\bf Summarizing topic phrases: }
After identifying topic phrases, we try to summarize the tweets corresponding to each of these topic phrases. Basically, we search the words present in topic phrases and retrieve those tweets that match. 
Finally, content words (nouns, numerals, verbs) based extractive summarization~\cite{rudra-cikm-2015} technique is applied over the retrieved tweets to generate a summary for each of the identified topics. 
Table~\ref{table:topic_summary} provides examples of identified topic phrases and their summaries.

\begin{table}[!htbp]
\center
\caption{{\bf Popular topic phrases and its summary for topics posted on first day of Nepal Earthquake (25th April)}}
\resizebox{\columnwidth}{!}{
\begin{tabular}{|p{0.3\columnwidth}|p{0.6\columnwidth}|}
\hline
{\bf Topic phrase} & {\bf Topic Summary} \\
\hline
communication cut & @AlwaysActions: China's \#Tibet severely affected by \#NepalEarthquake; houses collapsed, communications cut off \#Nepal\\
\hline
flight cancel & Flights to Kathmandu hit: Flight services to Kathmandu were today cancelled or put on ho; Kathmandu airport closed Saturday after a strong earthquake struck the country. All flights canceled.\\
\hline
\end{tabular}}
\label{table:topic_summary}
\vspace{-1mm}
\end{table}

The micro-level topics and summaries can be useful for various stakeholders in a disaster scenario. For instance, {\em Communication cut} can help government to plan, {\em airline held, flight cancel} can 
facilitate stranded foreigners to make proper departure planning while {\em medicine send} may enable the relief agency to connect supply to demand center.

~\\
\noindent {\bf Evaluating topic phrases: } 
To measure the accuracy of our proposed method for topic phrases identification, we check what fraction of nouns are correctly associated with the corresponding events. For this purpose, we compared the 
accuracy of our algorithm with a simple baseline algorithm in which nouns occurring within a window of 3 words on either side of the event were selected as being related to the event. Averaging over all 
the different classes (infrastructure, missing, shelter), the baseline algorithm obtains precision of 0.72, whereas our method obtains precision of 0.95.

Next, we evaluate the importance and utility of our identified topics. For this purpose, we performed user studies. For each day, we extract top ten topics based on our proposed methodology 
for each of the three classes. In a similar way, we identified ten topics using the LDA based topic summarization approach proposed by Arora~et~al~\cite{arora2008latent}. Each topic is represented by 
two words having the highest probability of belonging to that topic. We use a crowdsource based evaluation methodology to judge the utility of our topic based summarization approach over the 
LDA based technique. We asked five question to the workers on crowdflower as follows:
\begin{itemize}[leftmargin=*]
\item (Q1)~Relevance of the generated topics to the high-level category. (on a scale: from 1 (not related at all) to 5 (highly related)); 
\item (Q2)~Which method provides more situational awareness (M1 or M2); 
\item (Q3)~Which method shows less redundant topics (M1 or M2); 
\item (Q4)~Which method generates more semantically meaningful topic (M1 or M2); 
\item (Q5)~Usefulness of topic keywords for situational awareness (scale:1-5). 
\end{itemize}
By showing the top ten topics from both methods, we asked 15 different workers to answer each question.

\begin{figure}[!htbp]
\vspace{-2mm}
\center
\subfloat[{Relevance}]{\includegraphics[width=0.5\columnwidth]{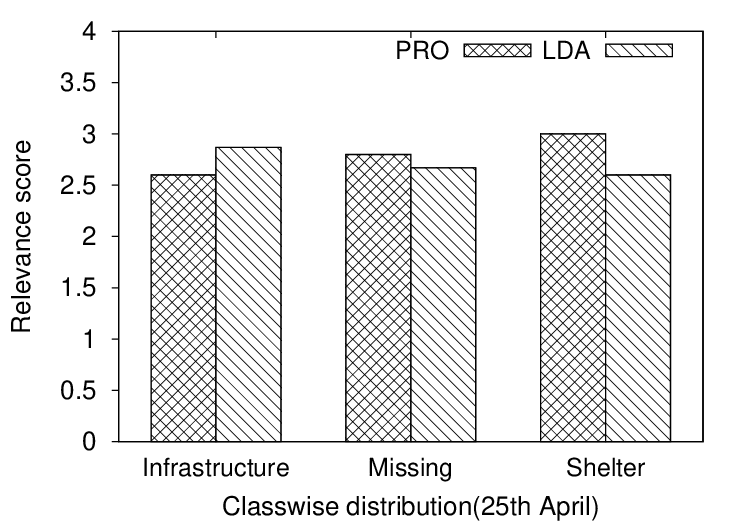}}
\hfil
\subfloat[{Situational awareness}]{\includegraphics[width=0.5\columnwidth]{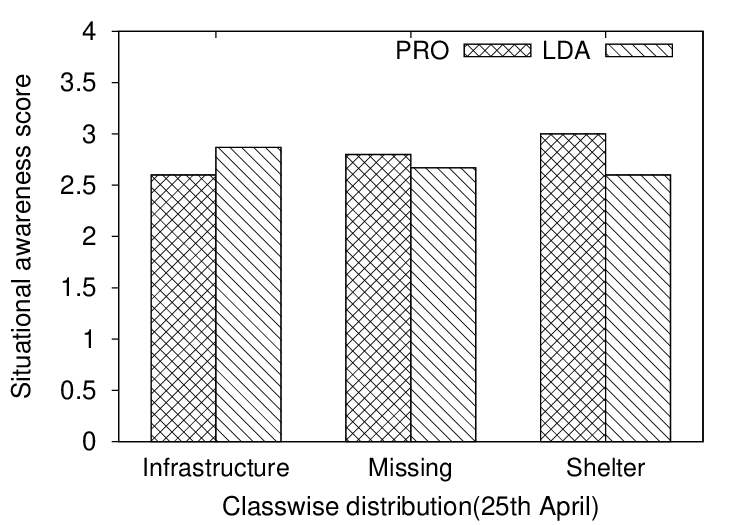}}
\caption{{\bf Results of the crowdsourcing based evaluation based on the relevance and situational awareness (25th April)}}
\label{fig:topic_crowd_score}
\vspace*{-3mm}
\end{figure}

With reference to the relevance to the high-level topics (i.e., Q1) and usefulness of topic keywords (i.e., Q5), out of nine cases, our method performs better than the baseline in six cases and
in rest of the three cases it is on par with the baseline. Figure~\ref{fig:topic_crowd_score} shows detailed ranking of users for 25th April for   Q1 and Q2. 
For questions 2, 3, 4, and 5, our method performs better than the baseline in six cases which demonstrates utility of our proposed topic detection scheme during crisis scenario.

\section{Conclusion}
\label{sec:conclu}

A large number of tweets are posted during disaster events. For better situational awareness, a concise, categorical as well as multi-faceted representation of the tweets is necessary. 
We presented a novel framework to summarize information in crisis-related tweets in two different forms: (a) general situation update summary and (b) specific flash point activity reports 
thus producing pointed information about a place and/or an event. To present such a diverse yet coherent picture, a deep understanding of the tweets posted during such scenario is necessary - we 
believe the series of innovations that have been undertaken in this work has been an outcome of thorough analysis of such tweets. We also have performed extensive evaluation using experts to determine 
the useful of our approach. In future, we will deploy the system so that it can be of help for any future disaster event.

\vspace{2mm}
\section*{Acknowledgement}
K. Rudra was supported by a fellowship from Tata Consultancy Services.

\balance
{
\bibliographystyle{abbrv}

}

\end{document}